\documentclass[aps,graphicx]{revtex4-2}

\usepackage{graphicx}
\usepackage{amsmath}
\usepackage{amsfonts}
\usepackage{amssymb}
\usepackage{subcaption}
\usepackage{epsfig}
\usepackage{epstopdf}
\usepackage{pstool}
\usepackage{soul}
\usepackage{psfrag}
\usepackage{babel}
\usepackage[pagewise,displaymath,mathlines]{lineno}
\usepackage{tikz}
\usepackage{xcolor}
\usepackage{mathtools}
\usepackage{hyperref}

\usepackage{array}
\usepackage{multirow}
\usepackage{setspace}
\newcommand{\nordita}{Nordic Institute for Theoretical Physics, Royal Institute of Technology \\and Stockholm University, Stockholm 10691, Sweden}

\begin{document}

\title{Stochastic Paleoclimatology: Modeling the EPICA Ice Core Climate Records}

\author{N. D. B. Keyes}
\email{ndbkeyes@aya.yale.edu}
\affiliation{Program in Applied Mathematics, Yale University, New Haven, Connecticut 06520, USA}
\affiliation{Department of Earth \& Planetary Sciences, Yale University, New Haven, Connecticut 06520, USA}

 \author{L. T. Giorgini}
\email{ludovico.giorgini@su.se}
\affiliation{\nordita}

\author{J.S. Wettlaufer}
\email{john.wettlaufer@yale.edu}
\affiliation{Program in Applied Mathematics, Yale University, New Haven, Connecticut 06520, USA}
\affiliation{Department of Earth \& Planetary Sciences, Yale University, New Haven, Connecticut 06520, USA}
\affiliation{\nordita}
\affiliation{Department of Physics, Yale University, New Haven, Connecticut 06520, USA}

\date{\today}

\begin{abstract}
We analyze and model the stochastic behavior of paleoclimate time series and assess the implications for the coupling of climate variables during the Pleistocene glacial cycles. We examine 800 kiloyears of carbon dioxide, methane, nitrous oxide and temperature proxy data from the EPICA Dome-C ice core, which are characterized by 100~ky glacial cycles overlain by fluctuations across a wide range of time scales. We quantify this behavior through multifractal time-weighted detrended fluctuation analysis, which distinguishes near-red-noise and white-noise behavior below and above the 100~ky glacial cycle respectively in all records. This allows us to model each time series as a one-dimensional periodic nonautonomous stochastic dynamical system, and assess the stability of physical processes and the fidelity of model-simulated time series. We extend this approach to a four-variable model with intervariable coupling terms, which we interpret in terms of \textcolor{black}{possible} interrelationships \textcolor{black}{among} the four time series. \textcolor{black}{Within the framework of our coupling coefficients, we find that carbon dioxide and temperature act to stabilize each other and methane and nitrous oxide, whereas the latter two destabilize each other and carbon dioxide and temperature. We also compute the response function for each pair of variables to assess the model performance by comparison to the data and confirm the model predictions regarding stability amongst variables.  Taken together, our results are consistent with glacial pacing dominated by carbon dioxide and temperature that is modulated by terrestrial biosphere feedbacks associated with methane and nitrous oxide emissions.}
\end{abstract}
%We draw conclusions about \textcolor{black}{potential} causal relationships in glacial transitions and the climate processes that may underlie these couplings, and highlight opportunities to further develop stochastic modeling approaches.

\maketitle

\begin{quotation}
Paleoclimate time series of greenhouse gases and temperature show Earth's periodic but noisy glacial transitions over the last 800,000 years, which are widely attributed to periodic changes in orbital forcing, but are still not well understood. Here, we apply a multifractal analysis method to four time series from the EPICA ice core to understand its colored-noise structure across timescales, showing that they are characterized by near-red noise on subglacial timescales and near-white noise on glacial timescales. Informed by this result, we model the time series as stochastic processes, first individually and then as a linearly coupled system, to extract stability and noise coefficients and assess interactions among the variables. \textcolor{black}{We examine the model coupling coefficients and compute the response function for each pair of variables, to reveal stabilizing and destabilizing relationships of varying strengths between them, suggesting potential causal relationships in climate transitions}. 
%The coefficients of our coupled model reveal that methane and nitrous oxide had a significant and destabilizing influence on carbon dioxide and temperature, showing that these greenhouse gases may have played an important role in leading glacial transitions.}
\end{quotation}

\section{Introduction}

\subsection{Background \& Motivation}

The Earth’s Quaternary glacial cycles are characterized by noisy processes with differing dynamics across timescales, but similar large-scale periodic behavior among different climate variables \textcolor{black}{that correspond to the glacial cycles of the Pleistocene. An area of particular interest in paleoclimate dynamics is the origin of the 100~ky cycle. The canonical explanation for glacial cycle pacing is the Milankovitch hypothesis, which attributes it to periodic changes in Earth’s orbital parameters \cite{hays_variations_1976}. Namely, because variations in Earth’s eccentricity, obliquity, and precession change the distance and angle of incident insolation to the planet’s surface over time, the resulting temperature changes are thought to drive the variations in greenhouse gas concentrations that are seen during glacial cycles. The Milankovitch cycle for eccentricity has an approximately 100 ky period, matching the glacial cycle periodicity.}

However, this hypothesis is the subject of great scrutiny, as evidence for it is typically based on pattern matching between the insolation and paleoclimate datasets. It is unclear why glacial cycles would be paced by eccentricity because it is the weakest of the Milankovitch cycles, as the hypothesis itself does not explain what kinds of mechanisms could amplify this small signal into one that dominates glacial pacing \cite{wunsch_quantitative_2004}. Furthermore, some examples of glacial termination data contradict the hypothesis, on the basis that changes in temperature precede their putative cause of changing insolation \cite{winograd_continuous_1992}, and hypothesis testing shows that the 100 ky eccentricity cycle specifically does not significantly influence glacial transitions \cite{huybers_obliquity_2005}. 
Indeed, a substantial challenge involves clearly identifying the internal climate mechanisms and feedbacks governing glacial cycles, and in particular the interactions between paleoclimate variables. \textcolor{black}{The community understands that many of the physical and chemical mechanisms that can facilitate these interactions, including the greenhouse effect, ocean carbon uptake, carbon rock weathering, soil nitrogen release, and permafrost melt \cite{dean_methane_2018,williams_carbon-cycle_2019,xu-ri_modelling_2012}, and seeks understanding of which processes may have dominated paleoclimate dynamics and hence may underlie the pace of glaciations}. \textcolor{black}{The issues were succinctly summarized by \citet{Berger:2003}:
\begin{quote}
{\em One of the most striking features of the 100 ky cycle is its pervasiveness, both geographically and within the various climatic subsystems. It dominates ice mass (and sea level), temperature, carbonate accumulation, upwelling, and carbon dioxide content of the atmosphere. This pervasiveness guarantees that (in the words of Laurent Labeyrie) ``everything is correlated with everything", which makes it difficult to deduce mechanisms from proxy records. }
\end{quote}
and by \citet{Imbrie:1993}:
\begin{quote}
{\em Dozens of explanations have been suggested (section 4). Some models explain the cycle as a free, self-sustaining oscillation with no Milankovitch forcing [e.g., Saltzman and Maasch, 1988]. In models of this type, the 100-ky cycle is forced by internal climate system processes so that its phase is arbitrary with respect to eccentricity. Other models explain the cycle as a nonlinear interaction between orbitally forced responses (in the 23- and 41-ky bands) and the internal dynamics of the atmosphere, oceans, ice sheets, and lithosphere [e.g., Maasch and Saltzman, 1990; Gallée et al., 1992]. In these, the phase of the 100-ky cycle is orbitally influenced.}
\end{quote}
For a recent review, the reader is referred to \citet{Ghil}.}

\textcolor{black}{Although our goal here is not to put forth a new theory for glacial pacing, we are interested in understanding the stochastic dynamics, noise characteristics and causal relationships among several key paleoclimate proxies that accompany glaciations.  To that end, the development of models that reproduce multiscale stochastic dynamics and elucidate causal interactions among climate processes are our focus. Many common statistical methods}, for example computing the covariance, can tell us the strength of the relationship between two variables, but cannot reveal the direction of cause and effect within that relationship, nor whether one process stabilizes or destabilizes another. \textcolor{black}{This problem can be addressed using a generalized Fluctuation-Dissipation Relation \cite{baldovin_extracting_2022}, which is able to identify causal links between the processes, but cannot reveal stabilizing and destabilizing relationships between them.} Global climate models simulate interactions in the climate system by numerically integrating conservation laws throughout the atmosphere and ocean and incorporating the influence of forcings and parameterization of relatively small-scale processes \cite{hansen_efficient_1983}. However, they often cannot reproduce the \textcolor{black}{variability, small-scale structure, and long duration} typical of climate time series due to the limited treatment of, and intermodel differences between, subgrid-scale processes \textcolor{black}{as well as the processing power needed to run such models over long time periods} \cite[e.g.,][]{stone_limitations_1990, Tapio:2020, Ma:2022, Alizadeh:2022}.

Paleoclimate analyses have examined causal relationships among paleoclimate data using various approaches, such as comparing prediction quality via convergent cross-mapping \cite{van_nes_causal_2015}, quantifying time lag between carbon dioxide and temperature at glacial transitions \cite{fischer_ice_1999}, calculating information flow among variables \cite{stips_causal_2016}, multivariate autoregressive modeling \cite{kaufmann_testing_2016} \textcolor{black}{or using the generalized Fluctuation-Dissipation Relation noted above \cite{baldovin_extracting_2022}}. \textcolor{black}{A multifractal method related to that described here was used by \citet{Shao:2016} to study the different scaling properties of interglacial and glacial climates using a wide range of data, including Antarctic and Greenland ice cores.  They found the Holocene record to be monofractal, and the glacial record to be multifractal, and concluded that the glacial climate has a longer persistence time and stronger nonlinearities.}

These approaches reach a variety of conclusions about the dominant causal direction among temperature and greenhouse gases and about the validity of the Milankovitch hypothesis \cite{Ghil}. This motivates new approaches of examining causal paleoclimate relationships. 

Our approach here is to use a stochastic data analysis and modeling method involving colored noise and non-autonomous stochastic dynamical systems theory. 
In the spirit of other stochastic dynamical systems theory approaches in climate science \cite{ghil_lucarini_2020, ghil_geophysical_2020, majda_models_1999},
we can characterize the random variability of climate processes that are not captured by deterministic models. 

\textcolor{black}{We first quantify the types of noise present in the time series \cite{epica_community_eight_2004} using a multifractal analysis method \cite{zhou_multifractal_2010} that allows us to identify the color of the noise in the record, and which colors characterize the dynamics over which time scales. This is essential for climate time series that exhibit both significant noise behavior and timescale separation, so that we can assess how the dynamics differ on shorter versus longer timescales.}

We then model paleoclimate time series as Ornstein-Uhlenbeck processes, consisting of periodic, nonautonomous Langevin equations that treat both the deterministic behavior and stochastic variability of the record. We apply these models to carbon dioxide, methane, nitrous oxide, and temperature proxy time series\textcolor{black}{, and assess their performance by computing the response function for each pair of variables.}

%\textcolor{green}{[(NASH) Ludovico - is there a better way to discuss the response function here?]}

Finally, we interpret the physical significance of the resulting model coefficients \textcolor{black}{and the response functions}, and pursue their implications for the interactions among these paleoclimate variables. With this in hand, we can simulate the time series and assess their fidelity relative to the original data through a variety of statistical metrics.

The structure of this paper is as follows. In Section II, we apply multifractal time-weighted detrended fluctuation analysis (MFTWDFA) to paleoclimate ice core records and quantify the nature of the fluctuations found therein. We introduce and apply our Ornstein-Uhlenbeck models in Section III and examine their properties and fidelity \textcolor{black}{ through statistical comparisons and computation of the response functions}. Having extracted the noise types present in these paleoclimate records, we reproduce the behavior and examine the causal relationships between proxies using simple stochastic models.  We conclude with a discussion of the implications of these analyses for the last 800,000 years of Earth's climate history. 
%The approach allows us to describe the causal relationships present in paleoclimate time series.

\begin{figure*}
    \centering
      \includegraphics[scale=0.22]{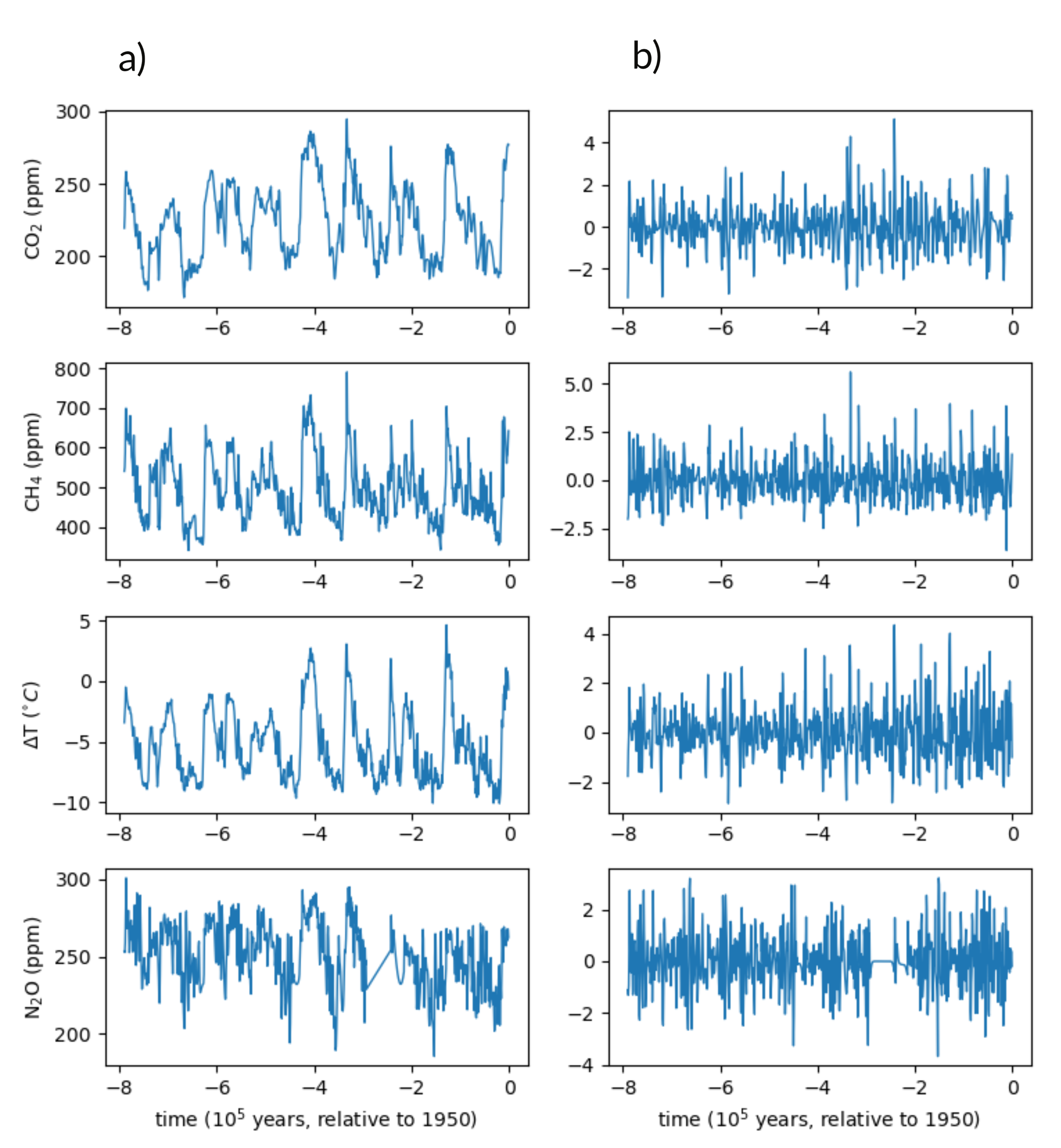}
     \caption{Carbon dioxide, methane, temperature, and nitrous oxide time series from the EPICA ice core record. (a) Original time series, (b) \textcolor{black}{normalized time series of the fluctuations relative to the slowly-varying mean}.}
    \label{fig:data}
\end{figure*}

\begin{figure*}
    \centering
       \includegraphics[scale=0.6]{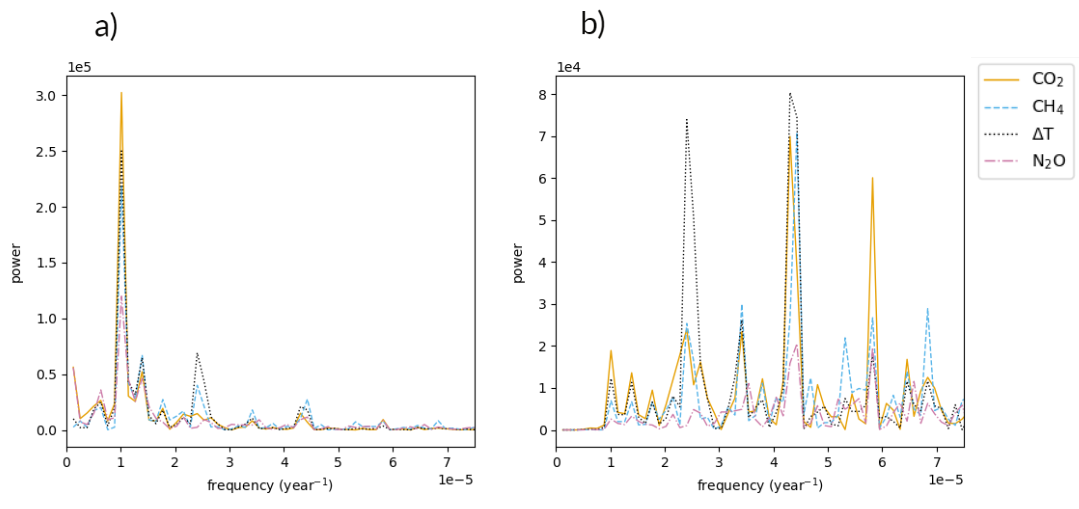}
    \caption{Power spectra of EPICA time series for (a) original data and (b) data fluctuations \textcolor{black}{relative to the slowly-varying mean behavior}.}
    \label{fig:spectra}
\end{figure*}

\subsection{The EPICA dataset}

\subsubsection{Background}

We analyze four 800 ky paleoclimate time series extracted from ice cores drilled at Dome C in Antarctica by the EPICA project. We examine records of carbon dioxide ($\text{CO}_2$), methane ($\text{CH}_4$), nitrous oxide ($\text{N}_2\text{O}$), and a proxy for temperature. Greenhouse gas concentrations are estimated by direct measurement of air bubbles trapped in the ice, while temperature is reconstructed from the deuterium proxy ($\delta D$) and presented as the change in temperature compared to the 1950 average global temperature. Direct measurements of nitrous oxide are supplemented by measurements of nitrous oxide artifacts where direct measurements were not possible. The four datasets use the EDC3 chronology, based on snow accumulation, flow modeling, and independent age markers, to estimate the correspondence between core depth and age \cite{parrenin_edc3_2007}.

As shown in Figure \ref{fig:data}, the 100 ky periodic glacial cycles are clearly observed in the time series. Additionally, however, the data also exhibit a complex, noisy structure across timescales. We first examine this structure by computing the frequency spectrum of all four time series, as shown in Figure \ref{fig:spectra}(a). \textcolor{black}{We see the strong $10^{-5}$ $\text{yr}^{-1}$ peak corresponding to  the period of 100 ky, the same period as the eccentricity cycle (lower-frequency peaks are not reliable because of the lack of data.), and, as noted above, is a feature that is the focus of a great deal of debate and research.}  

Furthermore, we observe \textcolor{black}{other relevant peaks} near $2.5 \times 10^{-5}$ and $4.25 \times 10^{-5}$ $\text{yr}^{-1}$, corresponding approximately to the 41 ky obliquity cycle and 23.5 ky net precession cycle resulting from the combination of axial and apsidal precession. \textcolor{black}{These peaks are typically attributed to the presence of external astronomical forcing in all the time series, which makes them highly correlated and, consequently, makes their causality relationships extremely difficult to unravel \cite{hays_variations_1976}. Therefore, we filtered this external forcing by subtracting from each time series a running average (see Section \ref{sec:data_prep} for more details). In Figure \ref{fig:spectra}(b) we show the frequency spectrum of each time series after applying the high-pass filter, and we observe that whereas the $10^{-5}$ $\text{yr}^{-1}$ peak is significantly reduced, the other two are increased.  Clearly, we have filtered (not expunged) the {\em time varying} external forcing and hence there remains an associated footprint in the time series, which must be taken into account.  This motivates building a non-autonomous stochastic system to model the filtered data and to examine causal relationships. }\\
%
%Therefore, even when filtered, the external forcing leaves a footprint in time series, which must be taken into account.  This motivates building a non-autonomous stochastic system to model the filtered data and to examine causal relationships. }

%\textcolor{green}{\emph{(John, do you have some physical reasons to explain how the external forcing influences the time series at short timescales?)}}\\
%
%\textcolor{blue}{\emph{The external forcing was not expunged, only filtered. Whence, it is reasonable to expect new resonances to arise.  However, without additional analysis, this is speculation and we ought not allow that to lead to more criticism.  The facts speak for themselves.}}\\

\subsubsection{Data preparation}\label{sec:data_prep}

In order to use the approach described above, we interpolate the time series to an evenly-spaced temporal resolution. We interpolate to match as closely as possible the lowest-resolution dataset \textcolor{black}{- nitrous oxide, with 912 points -} while also splitting the time domain into the 34 equal periods\textcolor{black}{, resulting in 25 points per period and a spacing of approximately 929 years between points. This constant spacing does mean that in the original time series, multiple points may be interpolated into some time gaps, but we confirmed that this is relatively uncommon and most time gaps in the original data are on the scale of this interpolation gap - the main exception being a large time gap in the nitrous oxide time series, which is an unavoidable limitation of the EPICA dataset.} For interpolation, we utilize the Akima method, which eliminates unrealistic overshoots introduced by other interpolation methods, such as the cubic spline \cite{akima_new_1970}\textcolor{black}{, particularly in the presence of large gaps in the dataset}. Moreover, other studies \cite{miller_testing_2019} have confirmed that interpolation generally does not impact results of statistical analysis. Due to the 20 ky gap in the nitrous oxide record from 260 to 240 ky, an artificial data point was added to the nitrous oxide time series at 250 ky using linear interpolation in that domain to better constrain Akima interpolation for our analysis. 

\textcolor{black}{After interpolating, we removed the slow-varying mean behavior, as we focus on modeling the smaller-scale fluctuations. We applied a Gaussian filter with a smoothing filter using a characteristic time-window three times the time increment used for interpolation. This approach resulted in an optimal filtering of slow fluctuations, while maintaining the fast fluctuations. We subtracted this mean behavior from the interpolated time series to obtain fluctuations around the mean. Subsequently, we normalized the fluctuation time series so that each has a standard deviation of unity, and thus can be modeled comparably. The distributions of the resulting fluctuation time series are nearly Gaussian, which supports our modeling approach described in  Section \ref{sec:mod}.}

\section{Multifractal Time-Weighted Detrended Fluctuation Analysis}

\subsection{Background}

We employ multifractal time-weighted detrended fluctuation analysis (MFTWDFA) \cite{zhou_multifractal_2010} to extract the scaling dynamics and fluctuation structure in the EPICA paleoclimate time series. This method quantifies the fluctuations in the time series around the mean behavior across timescales present in the data through the fluctuation function defined below. The approach enables us to draw conclusions about the dominant statistical fluctuations as a function of timescale. If the fluctuations in a time series are colored noise, the fluctuation function will scale exponentially over \textcolor{black}{increasingly large timescales}, and the particular value of the scaling exponent, referred to as the Hurst exponent, corresponds to the color. Therefore, a log-log plot of the fluctuation function is a straight line over the range of time in which the data exhibit a particular colored fluctuation behavior, and the slope of this line will be the corresponding Hurst exponent. \textcolor{black}{A power spectrum analysis could accomplish the same goal as MFTWDFA of quantifying colored noise, but the multifractal approach provides a clearer and more accurate description of the complex multiscale nature of the paleoclimate data and in particular the crossover times between regimes of noise behavior.}

MFTWDFA builds on other detrended fluctuation analysis methods such as MFDFA \cite{kantelhardt_multifractal_2002} by introducing a smoother computation of the mean behavior of the data on each timescale. 
In MFDFA a piecewise polynomial fit to the profile of the data is used. In MFTWDFA a time-weighted linear regression in a moving window provides a continuous estimate of the mean behavior at each timescale, leading to a fluctuation function that shows crossover times between noise regimes more clearly. Furthermore, MFTWDFA allows us to extract information about the nature of fluctuations at timescales up to $\frac{N}{2}$ for a dataset of length $N$, as opposed to $\frac{N}{4}$ in MFDFA.

We have used this MFTWDFA in previous work to extract the role of fluctuations in the dynamics of exoplanet detection, sea ice cover and global climate proxy data \cite{Agarwal:2017,Agarwal:2022,agarwal_trends_2012, moon_intrinsic_2018}. In order to make this presentation reasonably self-contained we outline the algorithm presently.

\subsection{Algorithm}

To calculate the fluctuation function, we construct a nonstationary profile $Y(i)$ of the original time series $X_i$, as
\begin{equation}
\label{eqn:profile}
Y(i) \equiv \sum_{k=1}^i \Big( X_k - \bar{X} \Big), \hspace{0.5cm} i=1,...,N.
\end{equation}
As noted above, in order to work with data evenly spaced in time, we interpolate $Y(i)$ using the modified Akima method.

Next, for each timescale $s$ in the data, the interpolated profile is detrended by removing behavior on timescales longer than that considered. This is done with a point-by-point approximation using weighted linear regression in a window of size $s$ around each point. The weights used in the local linear regression incorporate the intuition that points closer in time are more closely correlated than points farther away in time. Therefore, this continuously weighted fit smoothly captures the local mean and we determine the coefficients for the weighted fit, $\hat{\beta}$, at each point by solving
\begin{equation}
\label{eqn:wfit}
    (X^T W X) \hat{\beta} = X^T W y,
\end{equation}
where the elements of the weight matrix $W$ are defined as
\begin{equation}
    w_{ij} = \begin{cases}
    \Big( 1 - (\frac{i-j}{s})^2 \Big)^2 ,  & |i-j| \leq s \\
    0, & \text{otherwise.}
\end{cases}
\end{equation}

We then start at the beginning of the profile and split the data into intervals with an equal number of points, whose total time range corresponds to the timescale $s$. Accounting for the possibility that a portion of the profile remains, the same operation is reversed beginning at the end of the profile. In this manner, $2N_s$ segments are created, where $N_s = \text{int}(N/s)$ and $N$ is the number of points in the original series $X_i$.

For each timescale $s$, the variance of the data about the mean is computed up and down the profile, using
\begin{equation}
\label{eqn:var_up}
V(\nu,s) = \frac{1}{s} \sum_{i=1}^s \Big[ Y([\nu-1]s + i) - \hat{y}([\nu-1]s + i)\Big]^2 ,
\end{equation}
for $\nu = [1, N_s]$, and
\begin{equation}
\label{eqn:var_down}
V(\nu,s) = \frac{1}{s} \sum_{i=1}^s \Big[ Y(N - [\nu - N_s]s + i) - \hat{y}(N - [\nu - N_s]s + i) \Big]^2 ,
\end{equation}
for $\nu = [N_s+1,2N_s]$, where $\nu$ is the index of the moving time window of size $s$.

Finally, we obtain the fluctuation function, $F_q(s)$, as 
\begin{equation}
\label{eqn:fluct_func}
F_q(s) = \Big[ \frac{1}{2N_s}  \sum_{\nu=1}^{2N_s} \{ V(\nu,s)\}  ^{\frac{q}{2}} \Big]^{\frac{1}{q}},
\end{equation}
where $q$ denotes the statistical moment.

\subsection{Results: Data Analysis}

\begin{figure}[b]
    \centering
        \includegraphics[scale=0.63]{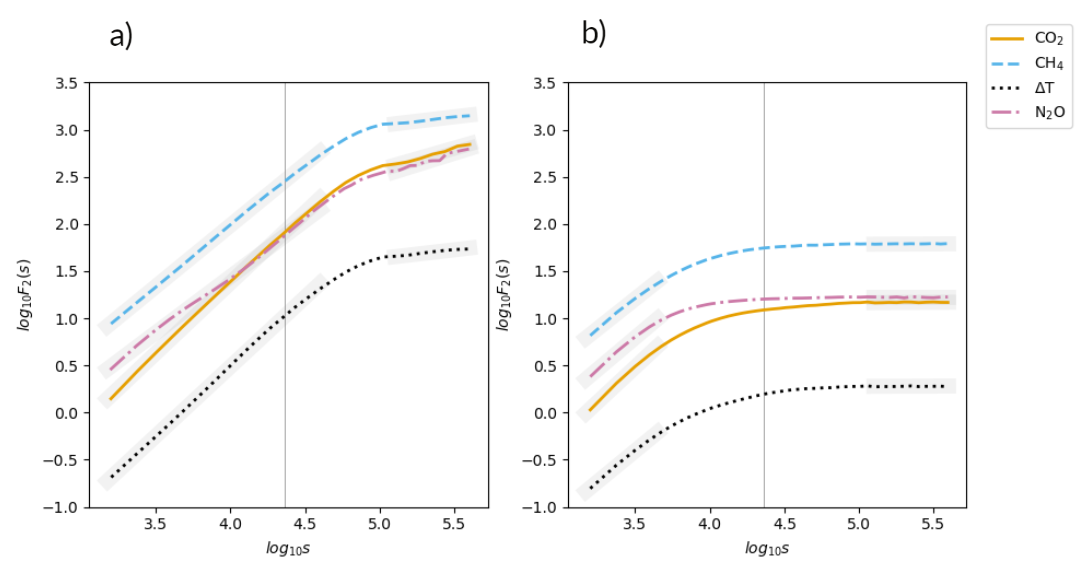}
    \caption{Logarithmic plots of the fluctuation functions for (a) original EPICA time series and (b) time series of fluctuations, with the slowly-varying behavior removed. Coarse wide gray lines show regressions fitted to approximately straight segments of the fluctuation function that correspond to distinct regimes of colored noise behavior, and vertical dotted grey lines show the 23 ky periodicity used in the later modeling section.}
    \label{fig:mftwdfa}
\end{figure}

\begin{table}

\centering
\renewcommand\arraystretch{1.5}{
a) \hspace{0.25cm}
\begin{tabular}{ l | l | l |}
  & $s < 10^{4.6}$ & $s > 10^{5.1}$ \\ 
 \hline
 \hline
 CO$_2$ & 1.50 & 0.46 \\  
 \hline
 CH$_4$ & 1.30 & 0.19 \\
 \hline
 $\Delta\mathrm{T}$& 1.48 & 0.18 \\
 \hline
  N$_2$O & 1.19 & 0.49 \\
  \hline
\end{tabular} \hspace{1cm} b) \hspace{0.25cm}
\begin{tabular}{ l | l | l |}
  & $s < 10^{3.6}$ & $s > 10^{5.1}$ \\ 
 \hline
 \hline
 CO$_2$ & 1.47 & 0.01 \\  
 \hline
 CH$_4$ & 1.26 & 0.01 \\
 \hline
 $\Delta\mathrm{T}$& 1.30 & 0.01 \\
 \hline
  N$_2$O & 1.34 & 0.00 \\
  \hline
\end{tabular}

}

\caption{Scaling exponent estimates from MFTWDFA fluctuation function slopes for the two colored-noise regimes, using $s < 10^{4.5}$ for the shorter-timescale regime and $s > 5.1$ for the longer-timescale regime. (a) Original EPICA time series, (b) forcing-converted time series, based on linear regression fits to the fluctuation function below and above the 100 ky glacial cycle crossover.}
\label{table:slopes}
\end{table}

For $q=2$ we fit straight line segments to the logarithmic plots of the fluctuation functions, which allows us to determine the Hurst exponents of the time series at different timescales. 
We use the second moment due to the simplicity of the correspondence between the Hurst exponent, $h(2)$, and the noise type \cite{kantelhardt_multifractal_2002}. \textcolor{black}{One can relate $h(2)$ to the slope of the power spectrum $\beta$ as $h(2)=(1+\beta)/2$. For a white noise process $\beta=0$, and hence $h(2)=1/2$. For a red-noise process $\beta=2$, and hence $h(2)=3/2$.  Thus, the varying slopes of the fluctuation function curves demonstrate the different dynamical processes operating on various time scales.}
%When $h(2)$ takes values between 0 and 1/2 the system is governed by anticorrelated stationary noise. When $h(2)$ = 1/2 the system is governed by uncorrelated white noise, becoming positively correlated as $h(2)$ increases up to 1, which characterized pink noise. Finally, as $h(2)$ increases from 1 to 3/2 the system is characterized by an anticorrelated nonstationary process, saturating to an uncorrelated red noise process at 3/2.

Ideally, to demonstrate robust scaling behavior, the straight-line slope segments should span as many orders of magnitude as possible. However, the length and resolution of the time series set upper and lower bounds for the timescales over which we can examine the noise behavior.

\textcolor{black}{By looking at Figure \ref{fig:mftwdfa}(a), we can see that all four of these climate variables are governed by similar stochastic dynamics below and above the glacial cycle timescale of 100 ky. Furthermore,} the fluctuation functions for the four original EPICA time series clearly show two distinct regimes of colored noise behavior. We fit straight lines to these two regions, $s=[3.2,4.6]$ for the original dataset and $s=[3.2,3.6]$ for the fluctuations, and $s=[5.1,5.6]$ for the long-timescale side, and find their slopes in order to quantify their noise types. In the time span between 1.5 and 40 ky ($10^{3.2}$ to $10^{4.6}$ years), the fits of the fluctuation functions give \textcolor{black}{ $h(2) \approx 3/2$, the Hurst exponent of a red-noise process.} In the time span between 125 and 400 ky ($10^{5.1}$ to $10^{5.6}$ years), we find
\textcolor{black}{ $h(2) \approx 1/2$, exhibiting white-noise behavior.} The fluctuation function structure for nitrous oxide differs slightly from the others at smaller timescales, with some subtler crossovers rather than the single slope seen in the other datasets. \textcolor{black}{However, for the purpose of our modeling, we are mainly concerned with the fluctuation function slopes for the data after the high-pass filter has removed the slowly-varying behavior.}

%glacial crossover and the associated timescale of the periodic terms in the model, and the smaller crossovers are smaller than that period. Therefore, we approximate the sub-glacial time scale nitrous oxide structure as a straight line and find $h(2)$ by fitting to the same sub-glacial timescale range as for the other processes.

%Examining the Hurst exponents in more detail shows that while the noise structures are similar, they vary slightly across all four climate variables. On glacial timescales, the carbon dioxide and nitrous oxide time series exhibit approximately white noise structure, with $h(2) \approx$ 1/2, whereas methane and temperature are much more anti-correlated, with $h(2) \approx$ 0.18. The sub-glacial time scale exponents for carbon dioxide and temperature are approximately 3/2, corresponding to near-pure red noise (Brownian motion or random walk). However, methane and nitrous oxide have $h(2) \approx$ 1.3, indicating a slight anti-correlated noise in those processes on shorter timescales. All four of these climate variables are governed by similar stochastic dynamics below and above the glacial cycle timescale of 100 ky, which is the key observation for our modeling study.

 \textcolor{black}{We then normalized the data and applied the high-pass filter as described in the previous section, after which we applied MFTWDFA analysis, which is shown in Figure \ref{fig:mftwdfa}(b).
%We performed the same MFTWDFA analysis on the data after \textcolor{black}{normalizing them and applying the high-pass filter, as described in the previous section, the results of which are shown in Figure \ref{fig:mftwdfa}(b). 
The glacial-scale slopes fall from $1/2$ to 0 and the crossover is shifted toward shorter timescales. A Hurst exponent  of zero indicates lack of scaling behavior on longer scales, and thus mean-reversion behavior of the time series.}  \textcolor{black}{ This can be attributed to the fact that the filtered time series describes a stochastic process that decays towards a constant position rather than a slowly time-varying signal. As the length of the smoothing windows is reduced, higher frequencies are removed from the data, leading to the observed shift in the transition point towards shorter time scales in the figure. The Hurst exponent for time scales below the length of the applied smoothing window of approximately 3 ky remained unaltered since the high-pass filter does not remove frequencies below the smoothing window. As a result, we can confidently use a multidimensional non-autonomous Ornstein-Uhlenbeck process to model the data on these time scales.}

\section{Stochastic models\label{sec:mod}}

\subsection{Background}

Climate time series can be modeled via simple stochastic processes if there is a clear separation between short and long timescales with distinct dynamics, and if the short-term processes can be modeled as random walks \cite{hasselmann_stochastic_1976}. Such a modeling approach is of interest because it incorporates the small-scale random fluctuations typical of climate processes \textcolor{black}{into a modeling framework that can be run over much longer time scales than can be achieved by current global climate models. We are also interested in the simplicity of such models, which allows us to determine the parameters of interest, such as stability and noise amplitude, analytically, and we can easily introduce coupling functions that can illuminate the nature of the interactions in the climate system. }

Our results from MFTWDFA justify the use of such a framework to model and analyze the EPICA paleoclimate time series. The difference in short-term and long-term dynamics for all four variables shows a clear separation of timescales between the sub-glacial and super-glacial periods.  \textcolor{black}{Moreover, even after filtering the long-term glacial cycle behavior, the short-term sub-glacial behavior is a nonstationary, approximately red noise, time series.}
%And, the short-term sub-glacial behavior is a nonstationary, walk-like time series that is close to red noise\textcolor{black}{, even after removing the slow glacial cycle behavior}.

%Because the short-term behavior is slightly anticorrelated rather than purely red in the methane and nitrous oxide time series, this is not a perfect approximation, but it is close enough that the model is still useful.

 \textcolor{black}{The stochastic model we employ extends the Ornstein-Uhlenbeck process to a non-autonomous periodic system with a separation of timescales. An Ornstein-Uhlenbeck process is the overdamped limit of the Langevin equation describing Brownian motion, when the particle experiences the restoring influence of a local quadratic potential.  Thus, the potential causes the dynamics to be mean-reverting. We can add to this canonical stationary process a longer-timescale forcing term to represent the slowly-varying mean behavior of the glacial cycles. Therefore, such a model can appropriately represent the way our paleoclimate time series fluctuates around this slowly varying mean behavior.}

 %is similar to a Wiener process, or Brownian random walk, but has an added deterministic drift term that constrains the noisy process with a quadratic potential, which causes it to be mean-reverting. We can add to this canonical stationary process a longer-timescale forcing term to represent the slowly-varying mean behavior of the glacial cycles. Thus, this model can appropriately represent the way our paleoclimate time series fluctuate around this mean behavior.}
 
 \textcolor{black}{Our model coefficients – the drift and the noise amplitude terms – are time-dependent and periodic. 
We seek to model the strong Milankovitch frequencies present in the paleoclimate time series, and this periodicity allows us to derive the model coefficients from the periodic statistics of the data. 
%We exploit time scale separation and periodicity to extract the stability, noise, and couplings present in the time series as follows. 
We choose the period of the model coefficients $a(k)$, $b(k)$, and $N(k)$, described below, based on the power spectra of our climate variables, Milankovitch cycle periodicities, and the timescale-separated noise structure revealed in MFTWDFA. We compare the power spectra (Figure \ref{fig:spectra}) of our four time series to identify a frequency peak, corresponding to a Milankovitch period, that is common across all four datasets. We also examine the MFTWDFA fluctuation functions (Figure \ref{fig:mftwdfa}) to find such a peak at a timescale small enough that the time series exhibit red noise dynamics.  The largest such frequency is approximately $4.25 \times 10^{-5}$ year$^{-1}$, or a period of about 23.5 ky, which corresponds to the Milankovitch cycle for the combined effects of axial and apsidal precession \cite{hays_variations_1976}.}

%Our model coefficients – the drift term and the noise amplitude term – are time-dependent and periodic. The paleoclimate time series have strong Milankovitch frequencies present in them that we would like to model, and this periodicity allows us to derive the model coefficients from the periodic statistics of the time series. Moreover, we are concerned with the shorter-term stability, noise, and couplings present in the time series, which are not affected by the longer-timescale behavior of the glacial cycles themselves because they occur at the shorter timescale under study. So, we neglect the long-term mean in our analysis.

%We choose the period of the model coefficients $a(k)$, $b(k)$, and $N(k)$ based on the power spectra of our climate variables, Milankovitch cycle periodicities, and the timescale-separated noise structure revealed in MFTWDFA. We compare the power spectra (Figure \ref{fig:spectra}) of our four time series to identify a frequency peak, corresponding to a Milankovitch period, that is common across all four datasets. We also examine the MFTWDFA fluctuation functions (Figure \ref{fig:mftwdfa}) to find such a peak at a timescale small enough that the time series behave like red noise as per the MFTWDFA fluctuation functions. The largest such frequency is approximately $4.25 \times 10^{-5}$ year$^{-1}$, or a period of about 23.5 ky, which corresponds to the Milankovitch cycle for the combined effects of axial and apsidal precession \cite{hays_variations_1976}.

\subsection{One-variable model}\label{sec:1D_model}

 \textcolor{black}{We begin with a one-dimensional non-autonomous Ornstein-Uhlenbeck model in order to reproduce the behavior of a single time series, which is based on \citet{moon_unified_2017}, but omits the long-term mean background behavior represented by their $f(\tau)$ term.  The non-autonomous model for the time series of the variable $\eta_i(t)$ is 
\begin{equation}
\label{eqn:1D_model}
\frac{d\eta_i}{dt} = a_i(t) \eta_i(t) + N_i(t)\xi_i(t). 
\end{equation}
The periodic deterministic term $a_i(t)$ drives mean-reverting drift, and represents the stability of the process: If $a_i(t)$ is negative (positive) the system is stable (unstable) and fluctuations decay (grow) in time. This stability is modulated by the noise amplitude, or noise intensity, given by $N_i(t)$, which is also periodic and deterministic. Finally, $\xi_i(t)$ is uncorrelated Gaussian white noise, resulting in a red-noise stochastic process, which models the behavior of the paleoclimate data. }
 
We solve for $a_i(t)$  and $N_i(t)$ using a modified version of the procedure described by \citet{moon_unified_2017} as follows. We consider a time series with $M$ periods and a resolution of $T$ points within each period of length $P$. Thus, the periodic function $a_i(t)$ in Eq. (\ref{eqn:1D_model}) is defined as $a_i(t)=a_i([t/\Delta t] \,\textrm{mod} \,T)$, where $[.]$ is the integer part and $\Delta t = P/T$. The same is the case for $N_i(t)$. We determine the $a_i(k)~\forall~ k \in [1,T]$ from the analytic solution of the model as
\begin{equation}
\label{eqn:a_coeff}
a_i(k) \approx - \frac{1}{\Delta t} \frac{S_i(k) - A_i(k)}{S_i(k)},
\end{equation}
where the approximate periodic variance and autocorrelation of the time series of the data $X_i$ are defined as
\begin{align}
S_i(k) &\equiv  \frac{1}{M-1} \sum_{j=1}^M X^{j T+k}_i X^{j T+k}_i \approx \langle (\eta_i(k))^2 \rangle, \qquad \textrm{and} \\
A_i(k) &\equiv \frac{1}{M-1} \sum_{j=1}^M X^{j T+k}_i X^{j T+k+1}_i \approx \langle \eta_i(k)\eta_i(k+1) \rangle, 
\end{align}
respectively. 

Finally, combining this formulation of $a_i(k)$ with the model Langevin equation, the expression for $N_i(k)$ is
\begin{equation}
{N_i(k)} = \sqrt{\frac{\langle y_i^2(k) \rangle}{\Delta t}},
\end{equation}
where 
\begin{equation}
%N_i(k)\Delta W 
y_i(k) \equiv \eta_i(k+1) - \eta_i(k) - a_i(k)\eta_i(k)\Delta t. 
\end{equation}
Step by step details are given in the Supplementary Information of \cite{moon_unified_2017}.  

\subsection{Four-variable model}

We extend this one-dimensional model in order to treat multiple time series together, introducing coupling terms that represent the influence that each time series has on the others. These coupling terms allow us to make first-order estimates of the primary direction of influence between time series variables, information beyond what measures like the covariance can provide. We develop a four-variable model based on \citet{moon_coupling_2019} to incorporate $\text{CO}_2$, $\text{CH}_4$, $\text{N}_2 \text{O}$, and temperature time series and their couplings to each other. We note that this type of system can be extended to an arbitrary number of variables.

The system of four equations is
\begin{equation}
\begin{split}
    \frac{\mathrm{d}\eta_i(t)}{\mathrm{d}t} &= a_i(t)\eta_i(t) + N_i(t)\xi_i(t) + \sum_{j\neq i} b_{ij}(t) \Big[ \eta_j(t) - \eta_i(t) \Big]
\end{split}
\label{eq:4d}
\end{equation}
 \textcolor{black}{where $\eta_i(t)$ is the $i$-th time series, $a_i(t)$ is the deterministic stability term, $N_i(t)$ is the noise amplitude term, and {$b_{ij}(k)$} is the  linearized diffusive coupling term representing influence of {$\eta_{j}(t)$} on $\eta_i(t)$, as is common across a wide variety of systems (see \citet{Othmer}, \citet{Levin}, \citet{Louis}, and \citet{Turing}, for just a few of many examples).}

It is significantly simpler to find the $a_i(k)$ and $b_{ij}(k)$ in this case, because we can solve four matrix systems for them directly. Each system is constructed by separately multiplying one of the model equations by each of the $\eta_{i}(t)$, and then taking the ensemble average, resulting in
\begin{equation}
\scalebox{0.85}{
$
\begin{bmatrix}
E_{ii}, & E_{ij} - E_{ii}, & E_{ik} - E_{ii}, & E_{il} - E_{ii} \\
E_{ij}, & E_{jj} - E_{ij}, &  E_{jk} - E_{ij}, & E_{jl} - E_{ij} \\
E_{ik}, & E_{jk} - E_{ik}, & E_{kk} - E_{ik}, & E_{kl} - E_{ik} \\
E_{il}, & E_{jl} - E_{il}, & E_{kl} - E_{il}, & E_{ll} - E_{il}
\end{bmatrix}
\times\begin{bmatrix}
a_i(t) \\
b_{ij}(t) \\
b_{ik}(t) \\
b_{il}(t)
\end{bmatrix}
=
\begin{bmatrix}
D_{ii} \\
D_{ij} \\
D_{ik} \\
D_{il}
\end{bmatrix} ,
$
}
\end{equation}
where $E_{xy} = \langle \eta_x(t) \eta_y(t) \rangle$ and $D_{xy} = \langle \frac{d\eta_x}{dt} \eta_y(t) \rangle$, which we solve for the coefficients.

Finally, to find the $N_i(k)$ we multiply each equation by its corresponding $\eta_i(t+\Delta t)$ and take the ensemble average to obtain
\begin{equation}
    \begin{split}
    N_{i}^2(t) &= \Big\langle \eta_{i}(t + \Delta t) \frac{d\eta_{i}}{dt} \Big\rangle - a_{i}(t)\Big\langle \eta_{i}(t) \eta_{i}(t + \Delta t) \Big\rangle  \\
    &- b_{ij}(t) \Big[ \langle \eta_{i}(t + \Delta t) \eta_{j}(t) \rangle - \langle \eta_{i}(t + \Delta t) \eta_{i}(t) \rangle \Big] \\
    &- b_{ik}(t) \Big[ \langle \eta_{i}(t + \Delta t) \eta_{k}(t) \rangle - \langle \eta_{i}(t + \Delta t) \eta_{i}(t) \rangle \Big].
    \end{split}
\end{equation}

\subsection{Results: Model\label{sec:model}}

We apply this modeling approach to the EPICA paleoclimate time series to derive and interpret the stability, coupling, and noise coefficients for each of the four variables in the coupled system. We quantify the model fidelity by using these coefficients to simulate artificial time series and compare them with the original time-series data.  

\subsubsection{Deterministic stability coefficients}

 \textcolor{black}{The deterministic stability in the one-variable model, Eq. \eqref{eqn:1D_model}, is controlled by the coefficient $a(t)$, and the deterministic net stability of the four-variable model, Eq. \eqref{eq:4d},  is controlled by  $a_{i,4D}$ (net) $= a_{i,4D} - \sum_{j} b_{ij}$.} We note that these coefficients are comparable across models in each variable, as we would expect since they are treating the same process. However, the four-variable model is slightly more negative across all four processes, showing that the couplings between the processes \textcolor{black}{enhance the overall stability.
In both models, methane and nitrous oxide are more stable than are carbon dioxide and temperature, and hence their deterministic drift drives them more strongly toward the long-term mean behavior.}

\subsubsection{Coupling coefficients  and noise amplitude coefficients\label{sec:coupling}}

\begin{figure*}[h!]
    \includegraphics[scale=0.6]{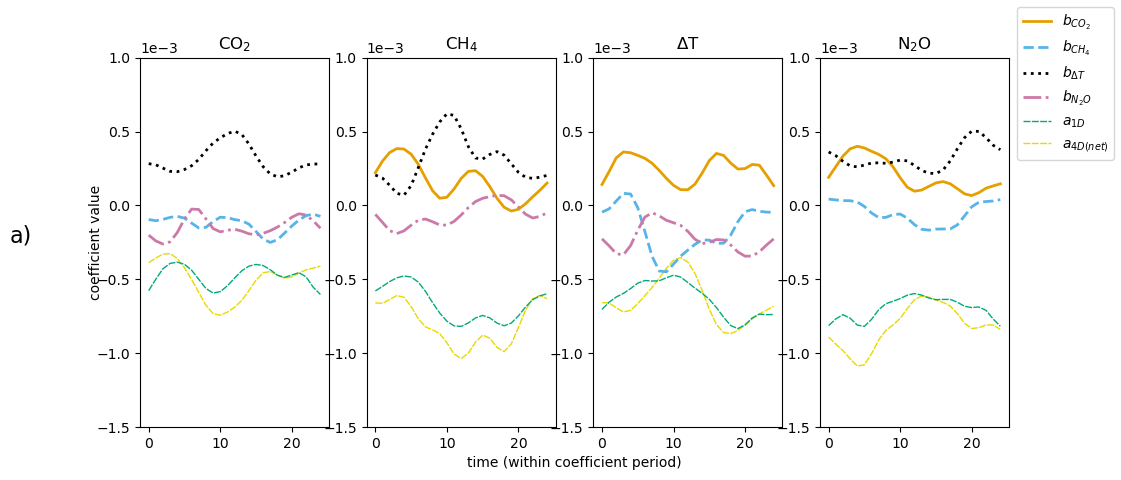}    
    \includegraphics[scale=0.52]{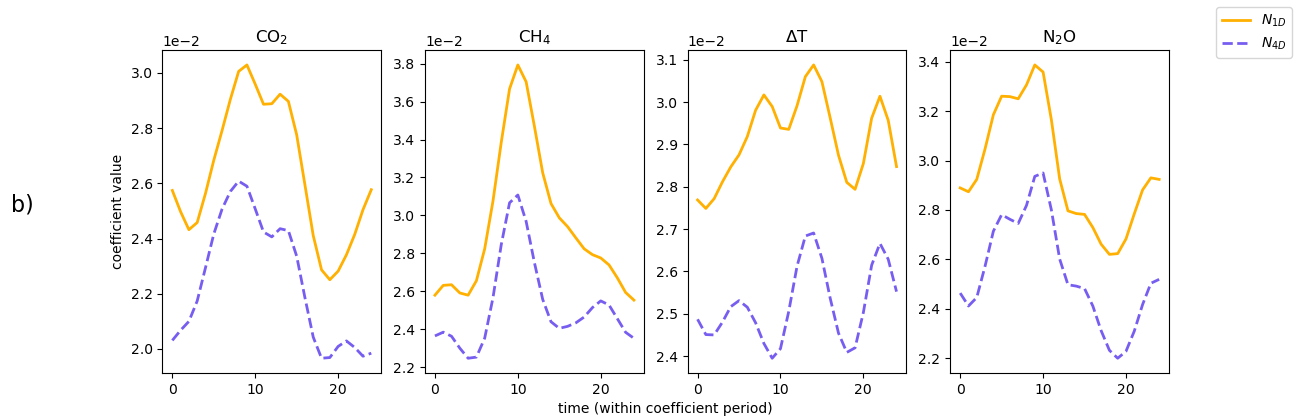} 
    \caption{\textcolor{black}{(a) Coupling coefficients for the four-dimensional model, the one-dimensional model stability and the four-dimensional model {\em net stability}, which is defined as $a_{i,4D}$ (net) $= a_{i,4D} - \sum_{j} b_{ij}$. (b) Noise amplitude coefficients for the one- and four-dimensional models.}}
    \label{fig:coeffs}
\end{figure*}

\textcolor{black}{The stability and coupling coefficients for the four-variable model are shown in Figures \ref{fig:coeffs}(a). The magnitude and sign of the coupling coefficients reflects the interactions between the processes.  For simplicity, consider only a two variable system, so that Eq. \eqref{eq:4d} becomes 
\begin{align}
\frac{\mathrm{d} \eta_1(t)}{\mathrm{~d} t} & =a_1(t) \eta_1(t)+N_1(t) \xi_1(t)+b_{12}(t)\left[\eta_2(t)-\eta_1(t)\right] \qquad \text{and} \nonumber \\
\frac{\mathrm{d} \eta_2(t)}{\mathrm{~d} t} & =a_2(t) \eta_2(t)+N_2(t) \xi_2(t)+b_{21}(t)\left[\eta_1(t)-\eta_2(t)\right], 
\label{eq:2d}
\end{align}
as in \citet{moon_coupling_2019}, where the coupling coefficient $b_{12}(t)$ ($b_{21}(t)$) represents the influence of variable 2 on variable 1 (variable 1 on variable 2).   
Therefore, the signs of the coupling coefficients characterize the direction of the influence that a pair of variables have upon each other, and the magnitude characterizes the strength of that interaction.  For example, when the coupling coefficient $b_{12}(t)$ is positive (negative) then the process represented by variable 2 suppresses (enhances) the growth of variable 1.  
Thus, in the canonical connotation of stability (instability) viz., the local in time decay (growth) of a variable, a positive (negative) coupling coefficient has a stabilizing (destabilizing) influence on the variables to which it is coupled.  Clearly, if $b_{12}(t)>0$ and $b_{21}(t)<0$ then variable 2 suppresses the growth of variable 1 and variable 1 enhances the growth of variable 2.}\footnote{\textcolor{black}{We note, however, that a different connotation of stability can be used \cite{moon_coupling_2019}.  Namely, in the context of the {\em longevity} of a climate variable, when $b_{12}(t)$ is negative (positive), and thereby provides a weak positive (negative) forcing to variable 1, we can say that variable 2 stabilizes (destabilizes) the {\em presence in the climate system} of variable 1.}}

\textcolor{black}{Finally, we note that \citet{Smale} showed that the deterministic form of Eq. \eqref{eq:2d} (i.e., $N_i(t)\xi_i(t) = 0$) is a structurally stable global oscillator, that is, apart from a closed set of measure zero, it has a nontrivial periodic attracting solution as $t \rightarrow \infty$.  The addition of the noise terms $N_i(t)\xi_i(t)$  in our model simply ``smears out'' the attracting solution to a degree that depends on the noise amplitude.  We return to this below. }

%\textcolor{blue}{(This paragraph might be put in a footnote, but it is how we discussed it in \cite{moon_coupling_2019}.)}\textcolor{black}{We note, however, that a different connotation of stability can be used.  Namely, in the context of the {\em longevity} of a climate variable, when $b_{12}(t)$ is negative (positive), and thereby provides a weak positive (negative) forcing to variable 1, we can say that variable 2 stabilizes (destabilizes) the {\em presence in the climate system} of variable 1.}

%The coefficients representing the influence of carbon dioxide and temperature on the other variables have consistently large positive values, whereas the coefficients representing the influence of methane and nitrous oxide are consistently small and negative, indicating that the former processes exert a stronger influence on the latter than vice versa

\textcolor{black}{We see in Figures \ref{fig:coeffs}(a) that, in the main, the coupling coefficients connecting carbon dioxide and temperature to each other or to methane or nitrous oxide have positive signs, and hence act to suppress the growth of these variables.  On the other hand, the coupling coefficients of methane or nitrous oxide are negative throughout, indicating that they act to enhance the growth of other variables.  The generally larger magnitudes of the coupling coefficients for carbon dioxide and temperature indicate their dominant control. Of course, the overall dynamics depends on all of the terms in Eq. \eqref{eq:4d}.}

%We note, however, that this is principally due to the relative abundances of $\mathrm{CH}_4$ and $\mathrm{CO}_2$ in the atmosphere rather than the intrinsic properties of the gases \cite{RayBook}.

 \textcolor{black}{The periodic behavior of the noise terms of the one- and four-variable models exhibit very similar dynamics across all variables.  For example, there is one significant peak in the middle of each period, with the exception of the two-peak structure of the one-variable model coefficient for temperature (Figure \ref{fig:coeffs}b). However, across all variables, the one-variable noise amplitude is consistently larger than the four-variable value. This is simply because the coupling terms in the latter provide additional sources of fluctuations, and hence each variable’s individual noise amplitude compensates by contributing a smaller amount of noise, thereby maintaining the same overall noise level between models.}

\begin{figure*}
    \includegraphics[scale=0.53]{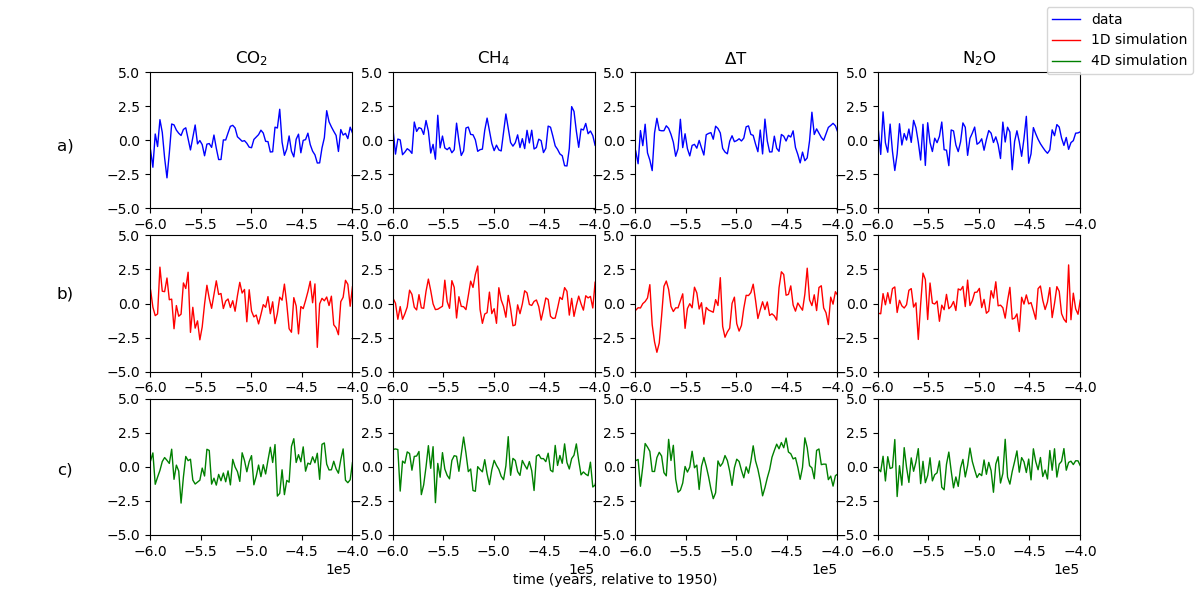} \hspace{0.4 cm}
    \caption{Comparison (from -600 ky to -400 ky) of (a) forcing data time series with simulated time series generated from (b) one-variable and (c) four-variable models.}
    \label{fig:sims}
\end{figure*}

\begin{figure*}
    \begin{center}
    \includegraphics[scale=0.42]{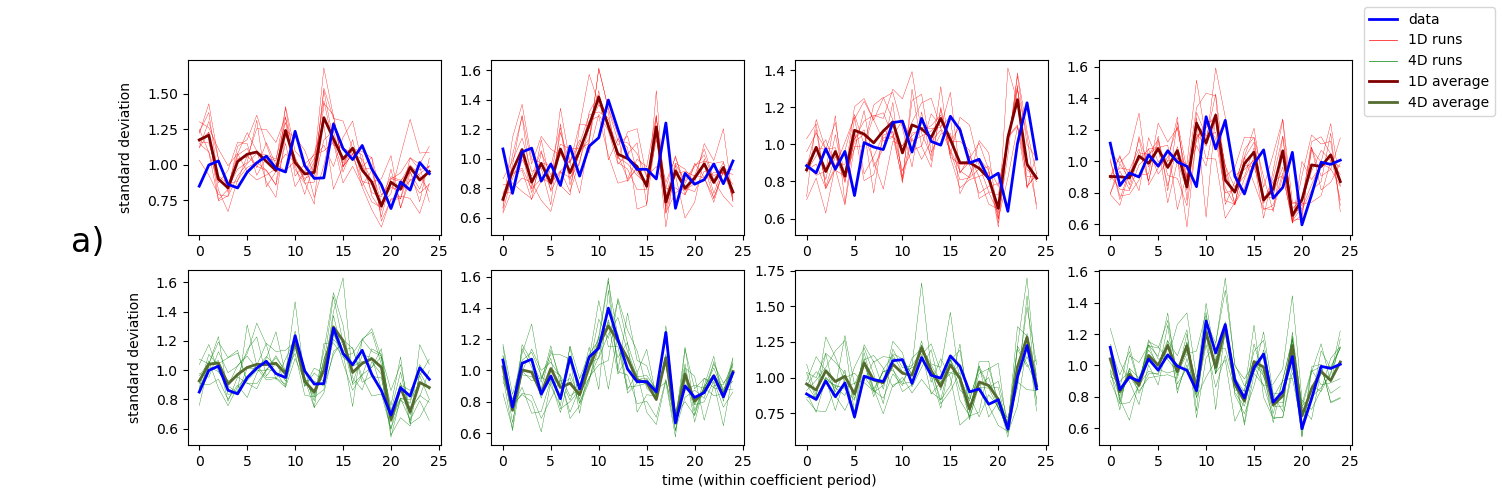}
    \includegraphics[scale=0.45]{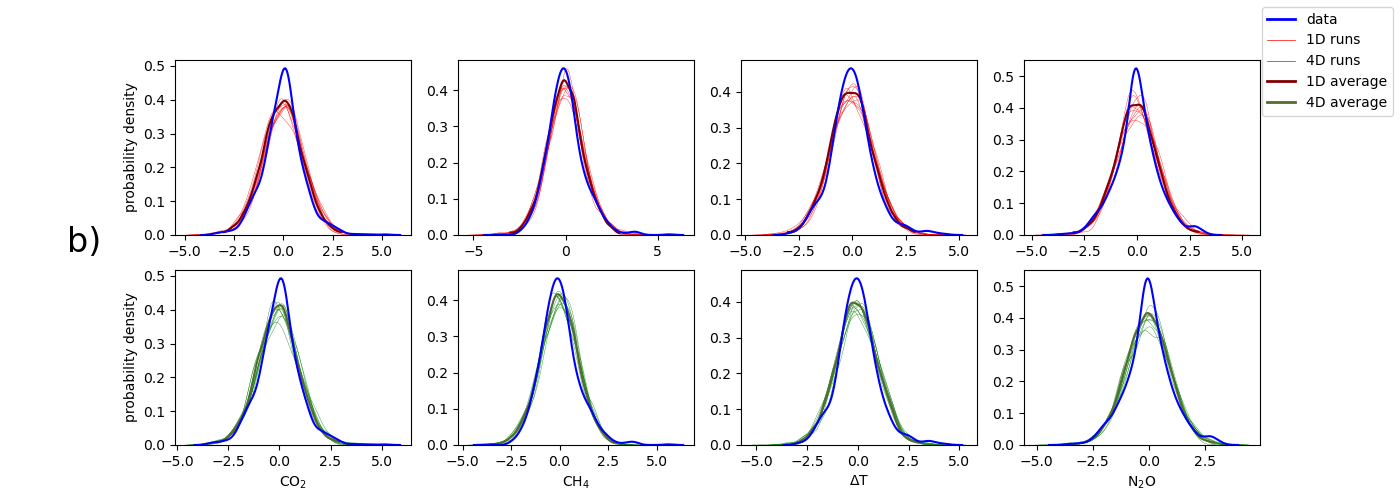}
    \includegraphics[scale=0.45]{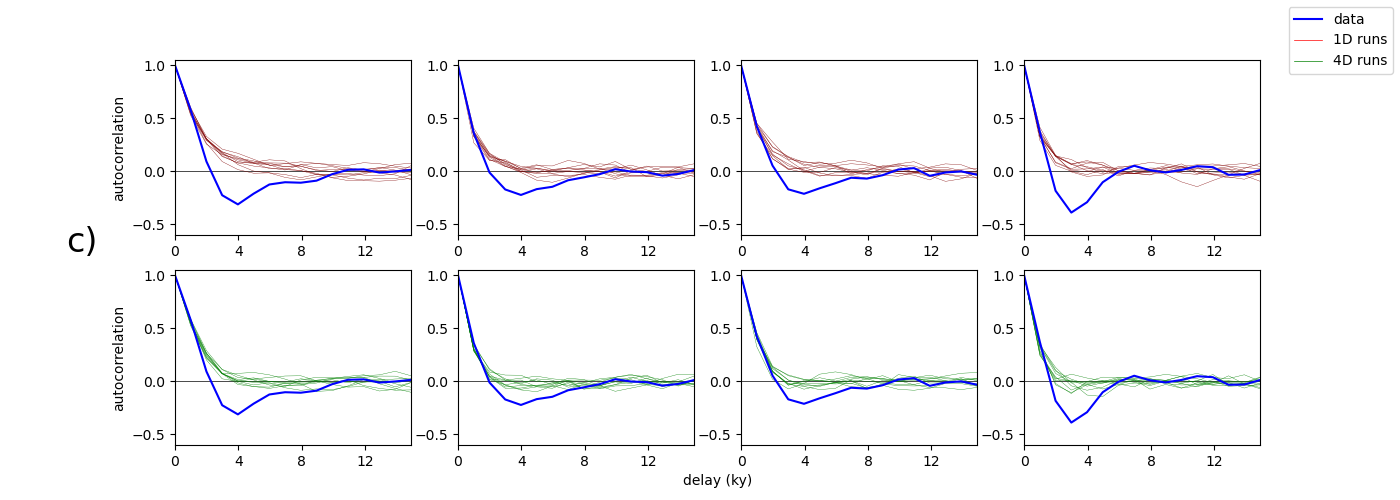}

    \end{center}
    \caption{Comparison of (a) periodic standard deviations, (b) probability density functions, and (c) autocorrelation functions between the data and one-variable (top rows) and four-variable (bottom rows) models. From left to right the columns are $\mathrm{CO}_2$, $\mathrm{CH}_4$, $\Delta\mathrm{T}$ and $\mathrm{NO}_2$.}
    \label{fig:compare}
\end{figure*}

\subsubsection{Model Interpretation\label{sec:interp}}

\textcolor{black}{The principal points at this juncture are as follows.  Across all proxies, the time average coupling coefficients for $\mathrm{CO}_2$ and $\Delta\mathrm{T}$ are positive, and hence their mutual interactions are stabilizing.  However, as seen in Fig. \ref{fig:coeffs}(a), depending on time, one can be larger than the other in an approximately periodic manner, so that the mutual stabilization is time dependent.  In contrast, the coupling coefficients for $\mathrm{CH}_4$ and $\mathrm{NO}_2$ are on average negative, but with a smaller magnitude than those for $\mathrm{CO}_2$ and $\Delta\mathrm{T}$. Thus, $\mathrm{CH}_4$ and $\mathrm{NO}_2$ have a weakly destabilizing effect. Finally, the deterministic stability coefficients are all negative.   }

\textcolor{black}{Clearly, the model captures the canonical strength of the $\mathrm{CO}_2$ and $\Delta\mathrm{T}$ covariation, and the positive feedback of that covariation on $\mathrm{CH}_4$ and $\mathrm{NO}_2$.  Contemporary studies show that, in general, warming-induced methane-climate feedbacks are positive, with the principal contributors being atmospheric methane lifetime and biogenic emissions from wetlands and permafrost \cite{methane}.  Such feedbacks are complicated by the fact that the terrestrial biosphere presently acts as a partial compensatory carbon sink of global emissions. 
%, which compensates up to $30 \%$ of global $\mathrm{CO}_2$ and $\mathrm{CH}_4$ emissions.  
Indeed, because the terrestrial biosphere is responsible for substantial fractions of $\mathrm{CH}_4$ and $\mathrm{NO}_2$ emissions, which increase under a warming climate \cite{Arneth:2010,Groenigen:2011}, these partially offset the cooling effect of the uptake of carbon by land \cite{Stocker}.
Moreover, because adding nitrous oxide (methane) is about 200 (20) times more effective at increasing global temperatures as adding equal amounts of carbon dioxide,  small fluctuations in the emissions of nitrous oxide and methane could be amplified into large effects on climate \cite[e.g.,][]{Khalil}.  We note, however, that this is principally due to their abundances in the atmosphere relative to $\mathrm{CO}_2$ rather than the intrinsic properties of the gases \cite{RayBook}.  Therefore, the modulation of the $\mathrm{CO}_2$ and $\Delta\mathrm{T}$ covariation by the warming of the terrestrial biosphere and the associated emission of $\mathrm{CH}_4$ and $\mathrm{NO}_2$, is consistent with the model presented here.  Namely, the relative magnitudes and signs of the coefficients are such that we view the $\mathrm{CO}_2$ and $\Delta\mathrm{T}$ covariation as the stochastic version of the \citet{Smale} global oscillator discussed in \S \ref{sec:coupling}, whose detailed evolution is influenced by the weakly destabilizing dynamics of $\mathrm{CH}_4$ and $\mathrm{NO}_2$. }

\subsubsection{Model fidelity}

We use the model coefficients computed from the EPICA time series and run our one- and four-variable models forward in time using a standard Euler method. This generates artificial time series with statistics and noise behavior that should match those of the original detrended time series. \textcolor{black}{Figure \ref{fig:sims} shows that both models reproduce the general appearance of the four time series.  Next we compare key statistical metrics to quantify how well our simulations reproduce different aspects of the proxy data.}

In Figure \ref{fig:compare}(a) we compare the periodic standard deviations of the data and models for each variable, and \textcolor{black}{find that the four-variable model is superior to the one-variable model in that it reproduces the overall magnitude and periodic shape of the standard deviation quite well for all of the time series}. The probability distribution functions also compare favorably, although we observe that the model fits a Gaussian distribution to the slightly non-Gaussian observations, as shown in Figure \ref{fig:compare}(b). \textcolor{black}{Thus, while both models reproduce the observational mean and standard deviation (within 3\% of the observational statistics in all cases), they do not reproduce the skewness and the kurtosis}.

In Figure \ref{fig:compare}(c) we compare the autocorrelation functions, which are less well reproduced than the other statistics. Whereas the one- and four-variable models both capture some of the oscillations in the autocorrelation function of the data, neither model reproduces the magnitude of the negative minimum of the data, nor the decay rate towards that minimum value. Here again, apart from some model approximations, which may not capture the full complexity of nonlinear processes in these paleoclimate time series, there may be many additional variables in the observed system that couple to those four in the observed record, but cannot be reflected in the four we treat in the model. \textcolor{black}{However, we note that the four-variable model reproduces the rate of decay in autocorrelation and some of the negative values  better than does the one-variable model, indicating an important role of the coupling coefficients.  Nonetheless, we view this behavior of the autocorrelation as a weakness in the predictive power of our approach.}

%\textcolor{blue}{Finally, in Figure \ref{fig:corr_rmse} we compare the correlation coefficient and the root mean squared error of the data to the four-variable model predictions, as a function of the time delay between the forecast initial value and the final prediction.  The figure illustrates a significant decay in the correlation coefficient over about a ky, which reflects the fact that a four-variable model does not capture the variability of a much higher dimensional system.}

%\textcolor{black}{Finally, in Figure \ref{fig:corr_rmse}, we compare the correlation (a) and root mean squared error (b) of the data and model forecast predictions, as a function of the prediction delay for the four-dimensional model. The figure illustrates the model's severely limited predictive capabilities, wherein the predictions become uncorrelated with the observations after a mere two-time-step delay. This result is expected, given that the model employed to represent a complex, nonlinear, and multidimensional system is low-dimensional and linear.}

\textcolor{black}{\subsubsection{Response functions}}

\textcolor{black}{Knowledge of the model coefficients allows us to construct the linear response matrix function, $R(\tau;t)$, which identifies the causal relations between each time series considered. For the model we study here, $R(\tau;t)$ can be written in terms of the time-dependent correlation matrix, $C(\tau;t)$, also called the persistence, according to 
\citet{baldovin_extracting_2022} as follows:
\begin{equation}
R(\tau;t) = C(\tau;t)C^{-1}(\tau;0),
\label{eq:response_num}
\end{equation}
where the matrix elements of the time-periodic correlation matrix are defined as
\begin{equation}
C_{ij}(\tau;t) = \frac{1}{M}\sum_{n=1}^{M}\eta_i(\tau+t+n T \Delta t)\eta_j(\tau+n T \Delta t),
\end{equation}
with $M$ and $T$ defined in Section \ref{sec:1D_model}. 
By expressing the time-dependent correlation function in terms of the model coefficients, we can write the following expression for the time-periodic response function
\begin{equation}
R(\tau;t) = \textrm{exp}\left[\int_\tau^{\tau+t}{\bf K}(t^\prime)\mathrm{d} t^\prime \right],
\label{eq:response_an}
\end{equation}
with $K_{ii}(t)=a_i(t)$ and $K_{ij}(t)=b_{ij}(t)$.
}

\begin{figure*}
    \begin{center}
   \includegraphics[scale=0.2]{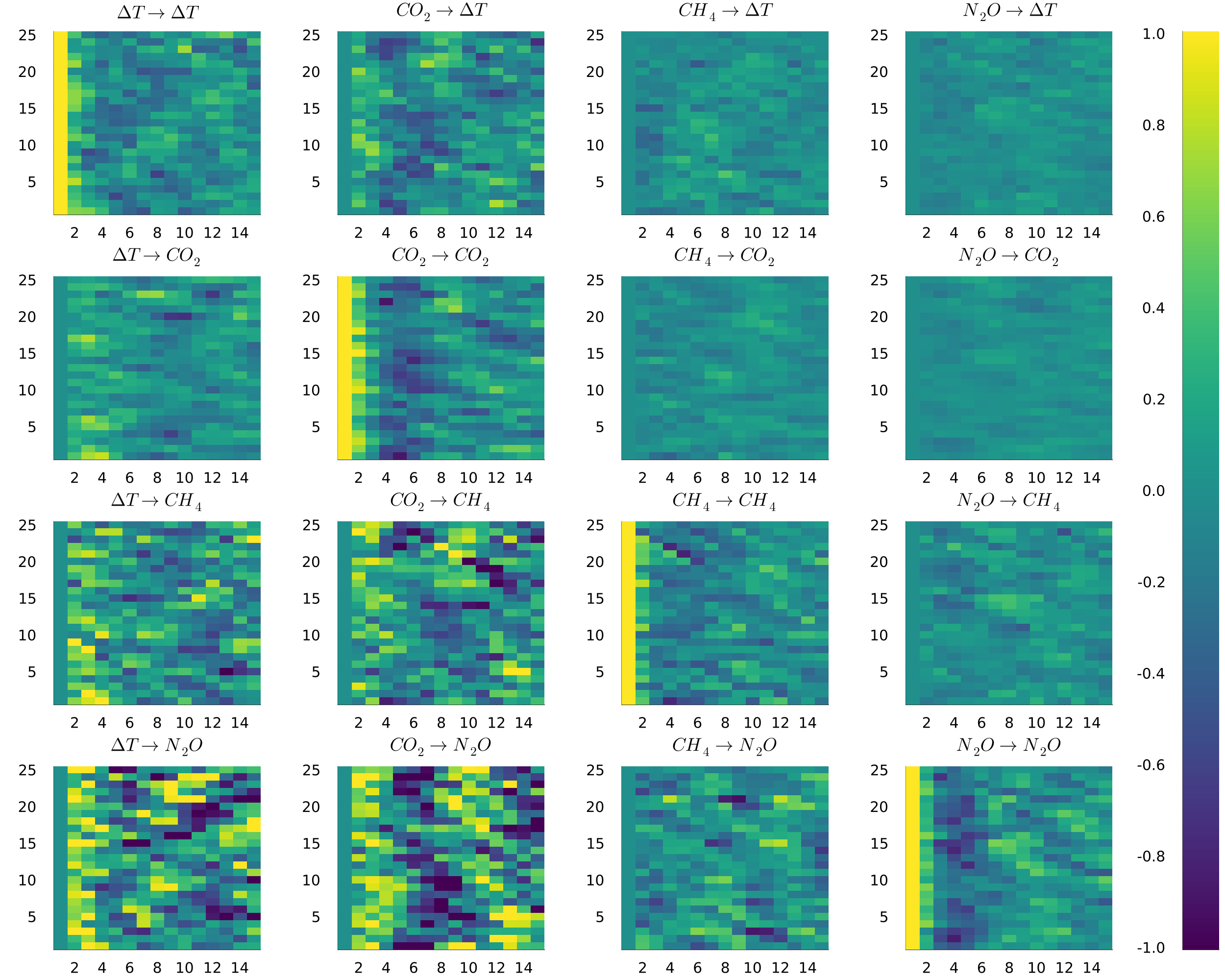}
    \end{center}
    \caption{Matrix elements of the time periodic response function obtained from the data plotted as a function of time. The response function is given by Eq. (\ref{eq:response_num}).}
    \label{fig:response_num}
\end{figure*}

\begin{figure*}
    \begin{center}
    \includegraphics[scale=0.20]{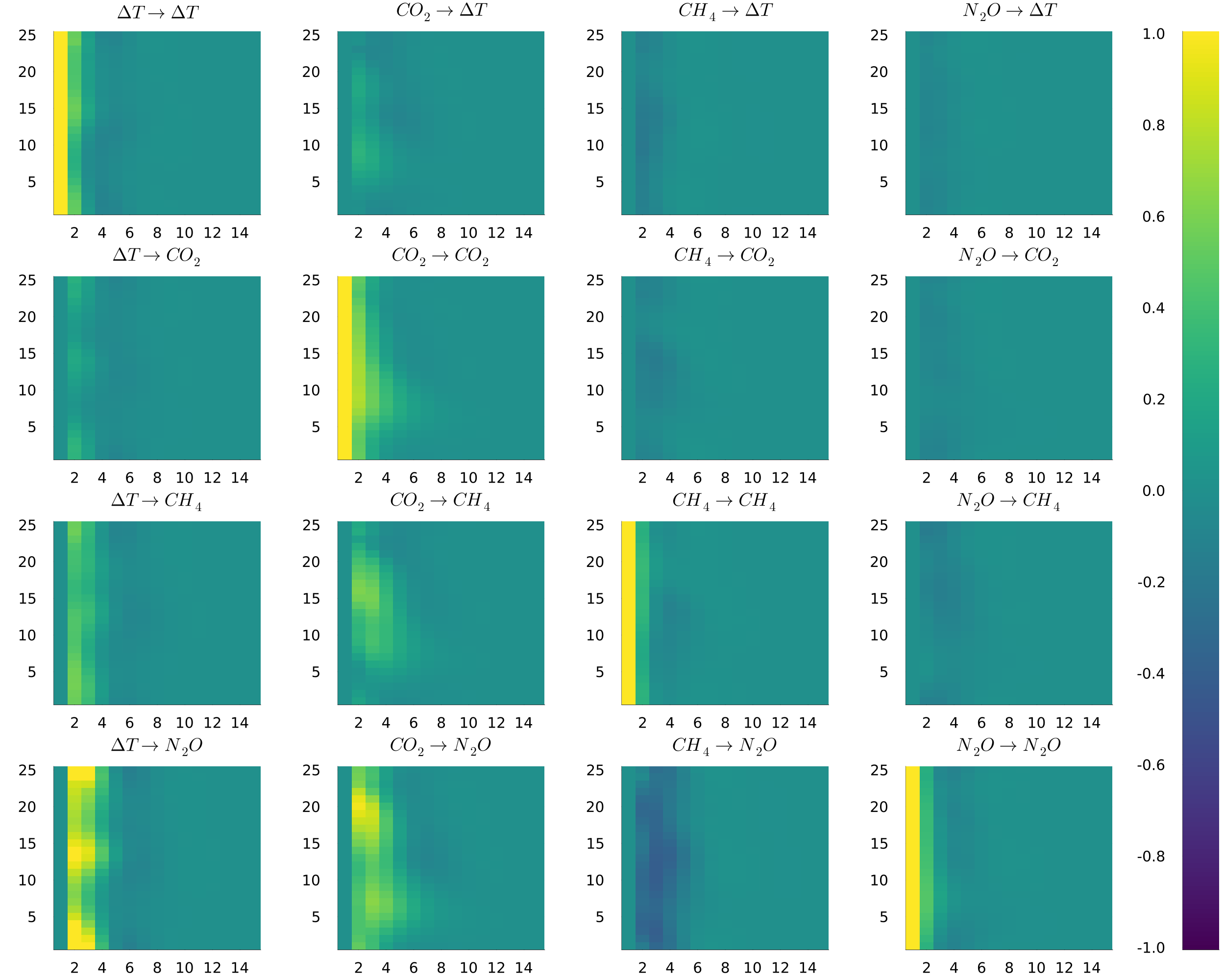}
    \end{center}
    \caption{Matrix elements of the time dependent response function obtained from the model coefficients plotted as a function of time. The response function is given by Eq. (\ref{eq:response_an}).}
    \label{fig:response_an}
\end{figure*}

\textcolor{black}{In Figure \ref{fig:response_num} we show the temporal behavior of the matrix elements of the response function constructed using the four different data sets, and in Figure \ref{fig:response_an} we show the matrix elements obtained from the model coefficients. Despite the inherent noise in Figure \ref{fig:response_num}, which results from averaging over only $M=34$ points (see Equation \ref{eq:response_num}), we observe qualitative agreement with Figure \ref{fig:response_an}. Specifically, we observe an overall stronger causal relationship from $\mathrm{CO}_2$  to $\Delta\mathrm{T}$ than from $\Delta\mathrm{T}$ to $\mathrm{CO}_2$, consistent with the findings of \citet{baldovin_extracting_2022}. However, the strength of these causal links varies throughout the period, with certain time windows displaying a stronger causal relationship from $\mathrm{CO}_2$ to $\Delta\mathrm{T}$, while others exhibit the reverse relationship. Furthermore, we observe that $\mathrm{CH}_4$  and $\mathrm{NO}_2$ have a negligible influence on $\Delta\mathrm{T}$ and $\mathrm{CO}_2$, while $\Delta\mathrm{T}$ and $\mathrm{CO}_2$ have a strong causal link to, and hence strong influence on, $\mathrm{CH}_4$  and $\mathrm{NO}_2$.  Importantly, this analysis of the response functions is consistent with the model interpretation discussed in \S \ref{sec:interp}.}

\section{Conclusion}

The Earth's paleoclimate underwent periodic but noisy 100 ky cycles of glaciation and deglaciation over the last 800 ky, which are clearly visible in time series data for carbon dioxide, methane, nitrous oxide, and temperature obtained from the EPICA ice core. We used a multifractal method to study these time series and extract the types of colored noise that characterize them \textcolor{black}{across scales, as well as the times at which there is a crossover between behaviors, in a more precise way than the usual spectral slope analysis allows}. This allowed us to adopt and extend previous non-autonomous stochastic models to represent each paleoclimate time series individually, and then as a coupled system, taking into account the time-dependent structure of their deterministic and stochastic dynamics.  

Our combined approach produces observationally consistent simple stochastic dynamical models. We extracted the timescale-separated colored noise regimes in the data, and computed and interpreted the stability, noise, and inter-variable couplings through non-autonomous Ornstein-Uhlenbeck models. \textcolor{black}{These coupling coefficients demonstrate the directionality and magnitude of the stabilizing effects of interactions between these climate variables, providing insight into the multiple time scale dynamics of the climate. }

%and potential counter-evidence to theory attributing glaciations to orbital cycles. 

%\textcolor{green}{(Nash) John - we need help translating our updated conclusions about stability relationships into conclusions about glacial cycles.}\\
%\textcolor{blue}{Here it is.}\\

\textcolor{black}{A central finding of our stochastic treatment is that carbon dioxide and temperature have stabilizing influences on each other and on 
methane and nitrous oxide, but the latter two have a weakly destabilizing influence on each other and on carbon dioxide and temperature. The strong co-variation between carbon dioxide and temperature has long been the signature of glacial cycles, but with the perennial question regarding which variable drives the other (see e.g., \citet{Kurt} and references therein).  Both the stochastic model coefficients and the response functions show this carbon dioxide and temperature  ``pulse'' of the climate system, but with a time-dependence of which one has  a controlling influence.  The weakly destabilizing influence of methane and nitrous oxide is due to the positive feedback--enhanced emissions--of the terrestrial biosphere to warming as discussed in \S \ref{sec:interp}.  \citet{Stocker} note that the contemporary terrestrial biosphere mitigates anthropogenic climate change by acting as a carbon sink, which compensates approximately $30 \%$ of global carbon dioxide emissions. Moreover, given the efficacy of methane and nitrous oxide as greenhouse gases, and the destabilizing influence we have identified our approach, it is clear that the carbon dioxide and temperature pulsing of glaciations is modulated by the terrestrial biosphere.  Keeping in mind that we have only modeled four proxies, we note that the asymmetry between stadials and interstadials, with the abrupt warming versus more gradual cooling, is consistent with our analysis.  The high latitude terrestrial biosphere is snow and ice covered during a stadial and the ice-albedo feedback exhibits hysteresis.  Thus, abrupt ice loss is accompanied by abrupt release of methane and nitrous oxide and thereby facilitates rapid warming.  During the warm interstadial slow terrestrial carbon uptake facilitates cooling until sufficient snow and ice cover suppresses terrestrial emissions driving the climate into a stadial. 
} 

%\textcolor{blue}{As noted above, one way to model the glacial cycle system is to use a double-well potential, switching between stable glaciated and deglaciated long-term states by passing through an unstable state. Therefore, coefficients with a stabilizing (destabilizing) influence will reduce (enhance) the likelihood that a process will drive a transition from one state to the other. 
%\\
%A central finding is that methane and nitrous oxide can have a destabilizing influence on the other processes, but carbon dioxide and temperature have stabilizing influences. This finding has several levels of interpretation.  The strong co-variation between carbon dioxide and temperature has long been associated with glacial cycles, but with the long standing question regarding which variable drives the other.  In this sense, our finding can be argued to show that these variables control or suppress the destabilizing influence of methane and nitrous oxide.  However, as noted above, one process can only drive a transition between states of another process if the coupling coefficient of the influencing process is negative, indicating that methane and nitrous oxide can drive glacial transitions reflected in carbon dioxide and temperature, but the converse is not possible.} 

On the one hand, \citet{Larsson:2014} and \citet{Persson:2019} used multivariate Granger causality tests in their analyses of the EPICA  ice core data to show that 
$\mathrm{CO}_2$, $\Delta\mathrm{T}$ and $\mathrm{CH}_4$  all ``Granger cause" each other.  Namely, their analysis strongly rejects the null hypothesis that any of these three variables does not cause the other.  
On the other hand, one of the important caveats discussed in \S \ref{sec:model} and mentioned throughout is our treatment of only four variables, which may themselves be coupled to others.  For example, the analysis of $\mathrm{CH}_4$ in the EPICA ice core data by \citet{Loulergue:2008} indicates that the connection between ice-sheet volume and Antarctic temperature and $\mathrm{CH}_4$ millennial variability, and the relationship proposed in the literature between them, fails to capture millennial $\mathrm{CH}_4$ events in the early glacial phases.  \textcolor{black}{Thus, the coupling between $\mathrm{CH}_4$ and other climate variables not treated here is non-trivial and not simply reflected in the coupling coefficients. Therefore, although the signature of glacial cycles is generally principally associated with the covariation of $\mathrm{CO}_2$ and $\Delta\mathrm{T}$, our results suggest that the interactive--coupled--role of other greenhouse gases is important in the timing of these cycles.  This realization is of course not new \cite{Arneth:2010,Groenigen:2011, Stocker}, but the point here is that it is cast in a framework that is much simpler to use than a comprehensive climate model.}

%\textcolor{blue}{Therefore, although the signature of glacial cycles is generally principally associated with the covariation of carbon dioxide and temperature, \textcolor{black}{our results suggest that taking into account} the role of other greenhouse gases may be more important than often assumed. For example, we know that although methane is a potent greenhouse gas, it occupies a relatively small wavelength window in the  outgoing long-wave flux of the planet. \st{However, according to our results, in one extreme methane may act as the dog of glacial transitions and carbon dioxide and temperature the tail. In the other extreme carbon dioxide and temperature may be the dog whose path fluctuates due to the influence of methane. Certainly, this speculation requires the scrutiny of further study.}}

\textcolor{black}{The approach described here constitutes a modest step in quantifying paleoclimate glacial dynamics using simple stochastic modeling techniques. Natural advances in the modeling framework would include, among others, nonlinear coupling between variables, nonlinear multiplicative and/or correlated noise. However, having examined only four paleoclimate observables, which may be coupled to many other variables, a clear next step is to introduce additional variables into our modeling framework. Clearly, this requires incorporation of more proxy variables thereby increasing the complexity of the coupled model, but such an approach is nonetheless vastly simpler than using comprehensive global climate models. Finally, because our approach reproduces key statistical dynamical quantities, it can in principle act as a constraint for comprehensive global climate models across a range of observationally accessible epochs. } 

\acknowledgements

N.D.B.K. and J.S.W. gratefully acknowledge support from Yale University.  L.T.G. and J.S.W. gratefully acknowledge support from the Swedish Research Council (Vetenskapsrådet) Grant No. 638-2013-9243. Nordita is partially supported by Nordforsk.

\section*{Author Declarations}
\subsection*{Conflict of Interest}

The authors have no conflicts to disclose.

%\bibliography{postgrad_paper.bib,postgrad_paper2.bib}

\providecommand{\noopsort}[1]{}\providecommand{\singleletter}[1]{#1}%\providecommand{\noopsort}[1]{}\providecommand{\singleletter}[1]{#1}%

\end{document}